**Attention acts to suppress goal-based conflict under high competition**

**Author names:** Claflin, O.[1,2]

Former Author Affiliations: [1]Departments of Psychiatry & Psychology, University of California, Los Angeles 90095, USA. [2]Department of Bioengineering, University of California, San Francisco, 94158, USA. [3]Departments of Physiology, Neurology and Psychiatry, and the Center for Integrative Neuroscience, University of California, San Francisco, 94158, USA.

Number of figures: 3; Number of tables: 3

Word count: 1350; Abstract: 88; References: 28

Acknowledgements: Thanks to Joseph Hsiao and Anh Tranh for collecting the data. Thanks to arxiv.org for a medium to post this manuscript on graduate work completed in Adam Gazzaley's lab at UCSF.

**Abstract**

**It is known that when multiple stimuli are present, top-down attention selectively enhances the neural signal in the visual cortex for task-relevant stimuli, but this has been tested only under conditions of minimal competition of visual attention. Here we show during high competition, that is, two stimuli in a shared receptive field possessing opposing modulatory goals, top-down attention suppresses both task-relevant and irrelevant neural signals within 100 ms of stimuli onset. This non-selective engagement of top-down attentional resources serves to reduce the feedforward signal representing irrelevant stimuli.**

It is well established that attention modulates visual processing in extrastriate cortex (Tsotsos et al., 1995). Strong evidence for the mutual competition theory, acting at the level of the receptive fields in the extrastriate cortex, suggests that local neuronal activity representing simultaneous stimuli result in suppression of these representations at the level of the receptive field (Moran and Desimone, 1985; Reynolds et al., 1999; Kastner and Ungerleider, 2001). This local suppression of local activity occurs automatically without top-down influences (Kastner, 1998). Many researchers have shown that top-down spatial attention serves to counteract the local suppressive influences on the stimuli of attentional interest present in the receptive field (Moran and Desimone, 1985; Luck et al., 1997; Kastner, 1998; Reynolds et al., 1999). This mechanism, acting at the level of the receptive field, spares the stimuli outside the field of directed spatial attention, leaving them suppressed, and thereby exerting selective enhancement on the stimuli of interest.

However, this mechanism of attention does not address the co-presence of task-relevant and a task-irrelevant stimuli in a shared spatially attended field. Long-range enhancement mechanisms would also enhance the irrelevant stimuli, acting as informational noise to the goal at hand, also present in the same receptive field, to be also amplified with the task-relevant stimuli. Otherwise, forced to act at the resolution of receptive fields, simultaneously presented objects with opposing behavioral relevance that share the same receptive fields may engage top-down noise-suppression mechanisms thereby suppressing the goal-relevant signal in order to suppress the goal-irrelevant signal. However, if these top-down mechanisms are able to act independently of receptive fields, then they will selectively act on these objects without modulatory limitation such that either the enhancement of relevant events or the suppression of irrelevant events, or both, will occur under these high competition settings (sharing space and timing).

Attentional selection behavior in the extrastriate cortex has been shown to be strongly dependent on attentional rivalry; namely, presentation containing simultaneously relevant and irrelevant stimuli have elicited new attentional behavior not previously observed (Zhang and Luck, 2009). Studies examining biased competition mechanisms have typically examined the suppressive interactions among irrelevant stimuli on each other (Beck and Kastner, 2009) but omit how competing goal-modulation in the same receptive field or goal-based attentional modulation is handled. Other studies examine simultaneous irrelevant and relevant stimuli sharing the same space and timing (Rutman et al., 2010; Gazzaley, 2011) but use discrete stimuli rather than continuous events. It is possible the ongoing representation result in more real-time stimuli processing mechanisms and, in fact, when used, has been observed to evoke novel attention behavior in studies that have used continuous events (Zhang and Luck, 2009).

Using electroencephalography (EEG) recordings and a continuous visuomotor task (modified from Anguera et al., 2013), we visualized stimulus competition affecting the earliest perceptual stream, as reflected by the P1 component, under different conditions of attentional rivalry (presence or absence of simultaneous relevant and irrelevant events; **Table 1**) and under different event timing overlaps (Event Onset Asynchrony (EOA); **Table 1**). A continuous visuomotor tracking task that required compensatory tracking adjustments was co-presented with a perceptual discrimination task that elicited an observable P1 component. Participants actively engaged with the tracking events (TID), the discrimination events (DIT), or both (MT). For our control condition, participants underwent an Ignore All (IA) condition where they simply attended to central fixation, while event onset asynchrony (EOA) between the two event types were jittered as in all the other conditions.

In line with previous findings (Hillyard et al., 1998; Beck and Kastner, 2009), P1 amplitude was larger for relevant events than irrelevant events (MT/DIT > TID/IA, $p \leq 0.028$; **Figure 3A**), and larger for isolated stimuli (DO) versus visually competitive stimuli (DO > MT/DIT/TID/IA, $p \leq 0.011$; **Figure 3A**). Additionally, P1 amplitude for behaviorally irrelevant events in our passive control condition (IA) did not show significant attenuation across EOAs ($p > 0.94$; **Table 3B**). Using this passive view as a baseline for all other conditions, we observed an impact of event onset asynchrony on conditions with an active task component (MT/DIT/TID; $p \leq 1.1 \times 10^{-3}$). Thus, using the P1 wave as a neural index, event onset asynchrony affected goal-based attention modulation of early sensory information within 100 ms of stimulus onset.

For non-simultaneous event onset (EOAs > 0 ms), irrelevant events remained at baseline levels of modulation (TID; $p \geq 0.3$) while relevant events demonstrated increased amplitude (MT, DIT; $p \leq 0.03$). For relevant events, a decline in this enhancing modulation was observed as event onset asynchrony increased (**Figure 3B**: MT: $F_{(1,19)} = 16.21$, $p = 7.2 \times 10^{-4}$; DIT: $F_{(1,19)} = 49.12$, $p = 1.13 \times 10^{-6}$). Thus the ability to selectively enhance relevant events during stimulus competition in the face of decreasing event recovery time diminished even when both events were relevant.

For conditions where competing attentional modulation was present (presence of relevant and irrelevant events; e.g. DIT, TID), simultaneous event onset resulted in a suppression of the P1 wave (**Figure 3B**: DIT, TID). Thus, in the case of simultaneous event onset (EOA 0 ms) and attentional rivalry, attention modulates the flow of early sensory information within 100 ms of stimulus onset by suppression of the event, regardless of its relevance. In contrast to previous competitive suppression findings, this suppression results in a signal strength below the passive view task (rather than just a relative suppression when compared to the single-task condition).

This observed decrease in P1 amplitude was not a bottom-up effect, as no decrease was observed in the signal strength in the passive condition sharing the same stimulus presentation (IA300/IA600 = IA0, $t_{(1,19)} \leq 0.253$, $p \geq 0.8$). Nor was the simple presence of top-down modulatory engagement explanatory as the dual-relevant condition did not exhibit this suppression (DIT0/TID0 < MT0, $t_{(1,19)} = 2.51$, $p \leq 0.05$). Rather, an interaction revealed the presence of competing top-down influences (Attentional Rivalry * EOA; ($F_{(2,38)} = 11.3$, $p = 1.4 \times 10^{-4}$)) explaining the pattern of P1 modulation in the setting of maximum stimulus competition (EOA 0 ms; **Table 2**: $p \leq 0.029$). However, as would be expected from the literature, post hoc t-tests revealed stimulus relevance was the explanatory factor affecting P1 modulation in less competitive settings (EOA 600 ms) (**Table 2**: $p \leq 0.006$), driving the stimulus relevance * EOA interaction ($F_{(2,38)} = 8.07$, $p = 1.1 \times 10^{-3}$).

These results suggest that attentional modulation responds to simultaneous goal rivalry between relevant and irrelevant events by suppressing early sensory representation for both events. Our experiment further demonstrates (1) that the competitive suppression between spatially-shared events is reflected in the visual signal, (2) that this competition is differentially alleviated by top-down enhancement of goal-relevant events, and (3) that this ability to selectively enhance linearly degrades with decreasing temporal spacing between events. This EOA-dependence of selective enhancement may represent a local neuronal competition effect in which local neurons representing these receptive fields become more sensitive to long-range enhancing projections with more recovery time between event onsets.

Previous research examining attentional modulation has shown attentional processing relies on general local suppression at the extrastriate cortex (Moran and Desimone, 1985; Reynolds et al., 1999; Beck and Kastner, 2005) accompanied by spatially-selective and feature-selective enhancement (Luck et al., 2000; Zhang and Luck, 2009), as well as a reliance on selective modulation at later stages of processing of object-relevant or irrelevant areas such as the FFA or PPA (Egner and Hirsch, 2005; Gazzaley et al., 2005). Considered together, these findings suggests that under conditions of high visual competition and opposing goal-relevance, top-down attention suppresses early stage signals to decrease feedforward signal from the irrelevant events and potentially rely on later processing stages of attentional selectivity for discrimination.

These findings support a goal-rivalry-dependent suppression of both relevant and irrelevant events during the earliest sensory processing stages in shared receptive fields under maximum temporal overlap. We hypothesize that this pattern of activity serves to reduce the feedforward signal representing irrelevant stimuli for further stages of processing, acting as an independent top-down suppression mechanism in the visual cortex. The observed suppression may represent an optimal strategy for the early selective attention mechanism when relevant and irrelevant events cannot be discriminated either spatially or temporally. In this case, reducing the early processing of the irrelevant event may be more advantageous compared to trying to enhance the relevant information along with unrelated noise. It may be that this mechanism of attention reflects a meta-monitoring of frontal processes that engage when lower-level features cannot selectively discriminate (Botvinick et al., 1999; Kerns, 2004; Carter and van Veen, 2007).

**Supplementary Methods, Results, Figures, Tables and Legends**

**Methods**

In the current study, we explored the hypothesis that selective attention mechanisms are limited in their ability to modulate under temporally, spatially and goal competitive conditions. We examine early signal modulation during a continuous paradigm with competing stimuli in a shared visual receptive field. We examined performance and neural measures while multitasking using a jittered design and concurrent electroencephalography (EEG) data collection to assess neural markers of perceptual processing. The paradigm involved variable event onset asynchronies (EOAs) between two tasks to assess the relationship between task timing and neural markers of early visual processing. We used a continuous visuomotor tracking task to generate continuous visual engagement and a forced choice perceptual discrimination task as a punctuated interruptor ('NeuroRacer': Anguera et al., 2013) to generate reliable event related potential (ERP) markers.

The paradigm has already been shown to generate interference costs on the behavioral and neural markers of the discrimination task (Anguera et al., 2013; Al-Hashimi et al., 2015), although the impact of EOA is unknown. We hypothesized that as the time between events in the visuomotor task (road turns) and events in the discrimination task (sign onset) was reduced, behavioral and early perceptual neural measures in the discrimination task would show greater interference costs, revealing how early stimulus processing is modulated by the complex demands of goal-related attentional modulation under increasingly temporally competitive conditions. Under these conditions we predicted that the effects of EOA and attentional goal on visual processing attentional modulation would interact to reveal limitations, or a new mode, of goal-related attentional processing.

**Materials and Methods**

*Participants.* Twenty healthy young, right-handed adults (mean age, 24.8 years; range, 20-29 years; 12 females) gave informed consent to participate in the study approved by the Committee on Human Research at the University of California in San Francisco. All participants had normal or corrected-to-normal vision as examined using a Snellen chart. Additionally, all participants were considered to be non-video game players, as defined by having less than 2 hours of any type of video-game usage per month in the past two years.

*Stimuli and experimental procedure.* Stimuli and tasks were presented using a custom designed video game ('NeuroRacer'; Anguera et al., 2013) on a Dell Optiplex GX620 with a 22" Mitsubishi Diamond Pro 2040U CRT monitor. Participants were seated with a chin rest in a dark room 80 cm from the monitor using a Logitech game controller to control tracking (left thumb) and responding to sign types (left & right index fingers). Each experimental run lasted 180 seconds with a few seconds of a self-paced break available to participants after every run.

The visuomotor tracking task (**Figure 1A**) is one of two tasks available in the 'NeuroRacer' environment (Anguera et al., 2013; Al-Hashimi et al., 2015), and involves keeping a car within a target box drawn on a continuously moving road (when the car was not within the target box, the fixation cross would shake to indicate poor performance). A pseudo-randomized, counterbalanced selection of road segments (that is, right/left turns & inclining/declining hills) formed the tracks, with turn/hill severity being either mild or severe. Transitions between road segments were treated as events in the visuomotor task, with each segment lasting 2000, 2500 or 3000 ms. This task involved continuous visuomotor tracking to an always inclining/declining and turning road. Tracking events involved transitions between these road segments where the inclination and/or turn direction and severity was changed suddenly. The point where the car crossed the union of these two road segments was treated as an event that required additional tracking correction to stay within the tracking zone.

The discrimination task (**Figure 1A**) involves responding to visual stimuli presented for 400 ms, less than two degrees above a fixation cross. Discrimination involved a 2-alternative forced choice task where subjects responded to green circles (33% frequency) with a right key press and all other colored shapes with a left key press.

Real-time feedback was indicated by a 100 ms color change of the fixation cross one second after stimulus presentation (green for correct, red for incorrect) for the sign task and by a shaking of the fixation cross when the car was outside the target-tracking zone.

*Thresholded performance.* Prior to the experimental runs, participants underwent an adaptive thresholding procedure to assess perceptual discrimination and visuomotor tracking abilities performed in isolation (**Figure 1A**). For discrimination, a staircase algorithm changed the time window allowed for a correct response for each 120 second run (48 signs) over nine runs. For the visuomotor tracking task, the speed of the road was thresholded with a similar staircase algorithm over twelve 60 second runs. The algorithm increased the difficulty of each task when performance level was over 80% on the previous run and decreased the difficulty when performance was less than 80% (for more details, see Anguera et al., 2013). Upon completion of each thresholding block, difficulty levels were interpolated for 80% performance to estimate an individual discrimination difficulty level and tracking difficulty level for that participant. The difficulty of the experimental tasks were set to these individualized levels for each participant so that individuals engage each condition in their own ability level following thresholding procedures, thus facilitating a fairer comparison across overall differences in perceptual discrimination abilities.

*Conditions.* Following the driving and sign thresholding procedures (**Figure 1A**), participants performed four perceptually-matched conditions and a fifth sign-only isolation condition randomly counterbalanced across participants (each condition performed three times in a pseudo-randomized fashion). Participants were cued to the upcoming condition before each run: 1: Multitasking (MT; Both tasks relevant); 2: Discrimination Ignore Tracking (DIT; Signs relevant); 3. Tracking Ignore Discrimination (TID; Road relevant); 4. Ignore All (IA; Neither task relevant, attend to fixation); 5. Discrimination Only (DO; Signs task present, on a black background). The road was present in all conditions (**Figure 1B**) except for Discrimination Only; signs were present for all conditions. These conditions included combinations of attend and ignore goals to the two tasks (**Figure 1D**), such that for the Discrimination Ignore Tracking and Ignore All conditions, the road task was instructed as irrelevant, driving feedback was turned off and the car was placed on 'auto pilot' for the duration of the run (**Figure 1B**). For the Tracking Ignore Discrimination and Ignore All condition, the sign task was made irrelevant through similar instruction and disabling of feedback and response mechanisms for the sign. In Multitasking condition, participants were told to respond to the signs as fast and accurately as possible and continue to drive as accurately as possible. Feedback was given at the end of each run as the proportion correct to all signs presented for the perceptual discrimination task and percent time spent in the tracking zone for the driving performance. Prior to the start of the subsequent run, participants were informed as to which condition would be engaged in next, and made aware of how many experimental runs were remaining.

*Event onset asynchrony.* Road events always preceded sign events (except in the case of 0 ms where they coincided) in the multitasking and distraction conditions. Event onset asynchronies of 0, 300 and 600 ms buffered the discrimination task from the road event (see **Figure 1C**) to observe the temporal impact on attentional modulation. This design was done not only for the multitasking condition (MT), but also for the conditions where both tracking and sign stimuli are present (TID, DIT, IA). Note that tracking events were not always preceded by sign events so that in this paradigm, tracking events an unreliable predictor for a discrimination event.

Participants were instructed to fixate at the center of the screen at all times, respond to signs as quickly as possible and keep the car centered in the tracking box at all times. Feedback was given at the end of each run as per their percentage correct on the discrimination task and percentage time spent within the tracking box. Correct responses to the appropriate signs within the thresholded response time window were categorized as hits; non-responses, late responses or mismatched key presses to stimuli were counted as incorrect for purposes of feedback. In the multitasking condition (MT), feedback was given for both tasks. Both tasks were equally emphasized in the instructions and briefing.

*Performance costs.* While both tasks were equally emphasized in instruction to the participants, we were focused on the performance effects on the discrimination task in our paradigm since it typically bears the brunt of multitasking and distraction costs (Anguera et al., 2013; Al-Hashimi et al., 2015). Driving-related metrics related for the visuomotor task are not discussed here. As an additional level of individualization beyond the individualized thresholding procedure, performance costs were compared for each participant against their undistracted single task performance. Distraction costs and multitasking costs were calculated by subtracting the participant's individual single task performance from their multitasking performance (MT - DO) and their distraction performance (DIT - DO) in order to individually baseline each participant's performance in these conditions. Response times (RTs) and accuracy were analyzed using ANOVAs, with a Greenhouse-Geisser correction when appropriate and *post hoc* t-test comparisons performed when statistically called for.

*EEG data acquisition.* Data were recorded during nine runs (three conditions, three runs each). Electrophysiological signals were recorded with a BioSemi ActiveTwo 64-channel EEG acquisition system in conjunction with BioSemi ActiView software (Cortech Solutions). Signal were amplified and digitized at 1024 Hz with a 24-bit resolution. All electrode offsets were maintained between $\pm 20$ mV.

*EEG data analysis.* Preprocessing and further ERP analyses was conducted in Analyzer 2.0 (Brain Vision, LLC). Raw EEG data were digitally re-referenced offline to the average of all electrodes. Eye artifacts were removed through independent component analyses by excluding components consistent with topographies for blinks and eye movements. Data were high-passed filtered at 1 Hz to exclude slow DC drifts.

An ERP analysis at posterior electrodes was conducted to assess markers of early visual processing for each sign presented. All ERP analyses were time-locked to the onset of each sign stimuli, yielding 216 epochs of data for each condition (and 72 epochs per EOA). Signals were averaged in 1000 ms epochs with a 200 ms prestimulus interval used as baseline. Epochs that exceeded a voltage threshold of $\pm$100 uV were rejected. The P100 and N200 component were evaluated at electrode clusters during the peak latency intervals of 100-180 ms for P100 and 140-260 ms for N200. To minimize electrode bias, activity at posterior electrodes was collapsed as follows: left-lateralized (P9, P7, PO7), right-lateralized (P1000, P8, PO8) and posterior midline (Pz, POz, Oz). Similar to the approach taken in other studies (Zanto et al., 2010; Mishra and Gazzaley, 2012), we used the greatest amplitude of the P100 and N200 in an electrode group when collapsed across all conditions to guide subsequent ERP analyses. This approach led to the posterior midline electrode group being selected for P100 interrogation and the right electrode group for N200 analysis.

To analyze the effects of Stimulus Relevance (Relevant, Irrelevant) and Attentional Rivalry (Present: [TID, DIT], Absent: [MT, IA]) (**Figure 1B**), across the three event onset asynchronies (0, 300, 600 ms), we used a 2x2x3 three-way ANOVA. ERP peak amplitudes were baselined against the control task amplitude (the IA peak was subtracted from peaks of the other conditions) for each participant to reduce intra-subject bias. The appropriate event onset asynchronies (IA0, IA300, IA600) were used to baseline all conditions with event onset asynchronies, and the subject's grand-averaged IA amplitude was used to baseline the amplitudes when analyzing the amplitudes of conditions collapsed across event onset asynchrony. ERP amplitude and latencies were analyzed using ANOVAs, with a Greenhouse-Geisser correction when appropriate and *post hoc* t-test comparisons performed when statistically called for.

*Current Source Density (CSD)*

With respect to the current study goals, it has been suggested that early visual ERPs like the P100 may benefit from current source density (CSD) filtering by removing volume conduction effects (Martinez et al., 2001). Resolving two discrete ERP events from stimuli that are presented closely in time has inherent technological issues (cf. Woldorff, 1993). CSD (Perrin et al., 1989) reduces redundant contributions due to volume conduction providing sharper topographies compared to those of scalp potentials (Kayser et al., 2006, 2007). Reducing volume conduction with CSD transformation may highlight local electrical activity at the expense of diminishing representation of distal activities (Cavanagh et al., 2010). Thus, we applied a CSD filter following preprocessing to reveal subtle visual cortical effects.

**Results - Behavioral data**

Performance costs

The influence of the EOA manipulation on multitasking costs (response times for the discrimination task during MT – response times for the discrimination task during DO) and distraction costs (response times for the discrimination task during DIT – response times for the discrimination task during DO) on the discrimination task was assessed using a 2 X 3 repeated-measures ANOVA with factors of Condition (MT, DIT) and EOA (0, 300, 600 ms). This analysis revealed a significant main effect of condition ($F_{(1,19)} = 95.7$, $p < 7.5 \times 10^{-9}$), event onset asynchrony ($F_{(2,38)} = 38.7$, $p < 6.8 \times 10^{-10}$), but no interaction ($F_{(1,19)} = 1.2$, $p = 0.33$). The condition main effect indicated that RT costs were greater during MT (48.4 ms ± 3.4) versus DIT (21.8 ms ± 1.2; $t_{(19)} = 3.17$, $p = 0.005$). In terms of the EOA effect, as the time between stimuli was reduced, costs became greater for both conditions (MT0 >MT300 >MT600: $t_{(19)} > 3.31$, $p < 0.004$ for each comparison; DIT0 >DIT300 >DIT600: $t_{(19)} \geq 1.99$, $p \leq 0.062$ for each comparison; **see Table 1 and Figure 2**). Reaction times were subtracted from individual's single-task performance in order to baseline the impact of the additional stimuli on the active task for each participant. Thus, the presence of a tracking event (road turn) (visuomotor task) relative to the time of sign onset (discrimination task) differentially impacted discrimination speed, regardless of whether the participant was actually tracking the road or the road was passively viewed.

A similar analysis of accuracy costs revealed a main effect of condition ($F_{(2,38)} = 8.9$, $p = 0.008$), but neither a significant main effect of EOA ($F_{(2,38)} = 2.5$, $p = 0.11$) nor an interaction ($F_{(2,38)} = 0.51$, $p = 0.61$). Follow-up analyses revealed participants to be less accurate when multitasking (84% ± 2) versus single-tasking with a road background (87% ± 2; $t_{(19)} = 3.35$, $p = 0.003$). Thus, the presence of an active secondary task (that is, visuomotor tracking) reduced discrimination accuracy compared to single-tasking, regardless of EOA.

**Neural data**

P100:

To examine the influence of attentional modulation on the early visual processing of the discrimination task, we examined ERP P100 amplitude and latency time-locked to the onset of the signs. For P100 peak amplitude, a 4 X 3 repeated-measures ANOVA with the factors of Condition (MT, DIT, TID, IA) and EOA (0 ms, 300 ms, and 600 ms) revealed a main effect of EOA ($F_{(2,38)}$ = 33.6, $p < 4x10^{-9}$) and a significant 2-way interaction ($F_{(4,76)}$ = 6.3, $p < .00019$), but no main effect of condition ($F_{(1,19)}$ = 2.11, $p = 0.14$). Pair-wise comparisons of EOA revealed an amplitude-reducing effect of decreasing EOA on P100 amplitudes for MT, DIT and TID, such that as sign presentation occurred closer to a road event (either a motoric road event (MT, TID) or a passive road event (DIT), the lower the P100 amplitude for that sign (DIT0 <DIT300 <DIT600: $t_{(19)}$>= 2.37, $p <= 0.029$ for each comparison; MT0 <MT300 =MT600: $t_{(19)}$>=2.38, $p <= 0.028$; TID0 <TID300 =TID600: $t_{(19)}$>= 3.25, $p <= 0.004$; **see Table 2 and Figure 3**). Additionally, pair-wise comparisons of conditions by EOA revealed MT and DIT to have greater P100 amplitudes at 600 milliseconds (**see Table 2 and Figure 3**).

P100 latency analysis revealed a main effect of EOA ($F_{(2,38)}$ = 7.09, $p = 2.4x10^{-3}$), but no main effect of condition ($F_{(3,57)}$ =0.788, $p = 0.51$) or interaction ($F_{(6,114)}$ = 1.61, $p = 0.152$). Post-hoc tests revealed that decreasing EOA, collapsed across condition, resulted in increased latency of P100 peaks (600: 151.7 ms; 300: 158.5 ms; 0: 164 ms; **see Table 3**).

The use of Ignore All condition to baseline the P100 ERPs when examining the impact of individual event onset asynchronies on peak amplitude and latency was appropriate as an ANOVA revealed no modulation on P100 amplitudes for the Ignore All condition ($F_{(2,38)}$ = 0.44, $p = 0.96$) nor for the IA latency ($F_{(2,38)}$ = 0.327, $p = 0.723$). One-way ANOVA's on unbaselined P100 amplitudes showed significant event onset asynchrony effects for all the other analyzed conditions (MT: $F_{(2,38)}$ = 8.26, $p = 0.001$; TID: $F_{(2,38)}$ = 9.4, $p = 0.004$; DIT: $F_{(2,38)}$ = 30.4, $p = 1.3x10^{-8}$).

To examine the effects of attentional rivalry and stimulus competition, a 2 X 2 X 3 repeated-measures ANOVA with the factors of Stimulus Relevance (**Relevant:** MT, DIT; **Irrelevant**: TID, IA), Attentional Rivalry (**Present**: DIT, TID; **Absent**: MT, IA) and EOA (**0** ms, **300** ms, and **600** ms) revealed no main effect of AR or SR ($F_{(1,19)} = 0.009$, $p = 0.927$; $F_{(1,19)} = 3.47$, $p = 0.078$) but a main effect of EOA ($F_{(2,19)} = 14.9$, $p = 1.6 \times 10^{-6}$). Significant 2-way interaction for AR*EOA ($F_{(2,38)} = 11.3$, $p = 1.4 \times 10^{-4}$), SR*EOA ($F_{(2,38)} = 8.07$, $p = 1.1 \times 10^{-3}$), but not for AR*SR ($F_{(1,19)} = 0.001$, $p = 0.97$). The 3-way interaction was not significant AR*SR*EOA ($F_{(2,38)} = 0.214$, $p = 0.81$). Exploration of these EOA interactions reveals a modulation pattern dependent on the individual relevance of that stimuli (Stimuli Relevance) at greater event onset asynchronies (600 ms); a modulation pattern that is impacted by the presence of opposing attentional goals (Attentional Rivalry) at event overlap (0 ms) (**see Table 2**).

If the greater multitasking amplitude over the single-tasking amplitudes in the competitive setting (MT0 > DIT0/TID0) was a result of cumulative enhancement of having two relevant events, we might expect to observe (DIT/TID > IA), which we do not ($p < 0.218$). Additionally, no detectable ERPs for road events were observed in our drive only (DO) condition ruling out an ERP contribution from the tracking task events. If selective enhancement was at work in our competitive setting (EOA 0 ms), we would observe a pattern where DIT > MT or DIT = MT but we actually observe the opposite (MT > SW, $p = 0.027$). Finally, we observe a significant ANOVA effect for attentional rivalry for 0 ms whereas we observe a significant main effect for stimulus relevance for 600 ms (Table 2A). In other words, in the setting where two events coincide, attentional rivalry ($p < 0.029$) rather than stimulus relevance ($p > 0.97$) becomes the determinant in the observed modulation patterns (**Table 2**).

N200:

We also assessed the N200 component, another early visual marker that has been associated with the identification or conscious processing of visual stimuli. Using the same analysis approach as for the P100, a main effect of EOA was seen for N200 amplitude ($F_{(2,38)} = 6.43$, $p = 0.0039$), but no main effect of condition ($F_{(4,76)} = 26.05$, $p = 0.68$) nor interaction ($F_{(2,38)} = 1.79$, $p = 0.18$) were present. Post-hoc tests revealed the attenuation of N200 amplitude with more temporally overlapping stimuli (**Table 1:** 0<300, 0<600, 300=600; 0 ms: -9.24 uV, 300 ms: -11.7 uV, 600 ms: -13.1 uV). N200 latency analyses revealed a main effect of condition ($F_{(1,19)} = 10.65$, $p = 0.0041$), but no main effect of EOA ($F_{(2,38)} = 1.35$, $p = 0.27$) or interaction ($F_{(2,38)} = 2.12$, $p = 0.13$). Post-hoc tests revealed N200 latency delay of multitasking (MT>DIT, **Table 1:** $t_{(19)} = 2.45$, $p = 0.024$; MT: 219.1ms, DIT:

207.9ms). Thus, increasing temporal overlap resulted in N200 amplitude attenuation, but no difference between conditions, whereas the addition of active tracking resulted in a non-EOA-dependent N200 latency delay.

*Eye Movement.* To ensure that the observed effects were not due to eye movement, electrooculographic data were analyzed (Anguera et al., 2013). Vertical (VEOG = FP2 – IEOG) and horizontal (HEOG = REOG – LEOG) difference waves were calculated from the raw data and baseline corrected to the mean prestimulus activity. The magnitude of eye movement was computed as follows: $(VEOG^2 - HEOG^2)/2$. The variance in the magnitude of eye movement was computed across trials at each time point between 0 and 200 ms post stimulus onset, which encompasses the P100 and N200, the ERP peaks of interest. The variance was compared between EOAs of 0, 300 and 600 via the two-tailed paired t-test. Uncorrected for multiple comparisons, no effects were observed at any time point tested regardless of EOA ($p > 0.05$), indicating that effects observed in the ERP are not due to eye movements.

**Legends**

Table 1. Multitasking and Distracted-Tasking Response Costs for the Multitasking (MT) and Discrimination Ignore Tracking (TID) conditions. The response costs were calculated as individual response time differences for each Event Onset Asynchrony (EOA) by subtracting the Discrimination Only (DO) response time (MT RT – DO RT; DIT RT – DO RT for both conditions respectively). P-values are reported for the individual Event Onset Asynchrony within and between conditions.

Table 2. Summary of the 2-way ANOVA analyses for P100 amplitude for the high temporal competition setting (EOA **0 ms**) and for the low temporal competition setting (EOA **600 ms**). A 2x2 ANOVA explored the effects of Attentional Rivalry (**Present** Competing Attentional Modulation: DIT, TID; **Absent** Competing Attentional Modulation: IA, MT) against Stimulus Relevance (Relevant sign events: MT, DIT; Irrelevant sign events: DI, IA). As expected, P100 amplitude for signs shows a main effect for stimulus relevance (MT, DIT > DIT, IA) in the low temporal competition setting. However, under high temporal competition, P100 amplitude is instead driven by the presence of top-down attentional rivalry rather than goal relevance. Post-hoc t-tests demonstrating the interaction across time and the four conditions are shown.

Table 3. Event-Related Potentials (ERPs) for sign events during the Multitasking (MT), Sign Ignore Road (DIT), Ignore All (IA), and Tracking Ignore Discrimination (TID) conditions. P-values are reported for ANOVA interactions, main effects and t-tests involving both EOA and condition for P100/N200 latency and amplitude. As seen, only the P100 amplitude demonstrated an interaction of attentional modulation and event onset asynchrony. B. P100 Amplitude Interaction Analysis for all Conditions. P-values are reported for one-way ANOVAs, for each condition.

Figure 1. NeuroRacer Tasks and Experimental Design. A. Screenshots of thresholding conditions. The Tracking Only and Discrimination Only conditions were used to establish titrated 'driving' and 'discrimination' levels in the setting of no distraction for each participant. These difficulty levels were used in the subsequent experimental conditions. B. Perceptually-matched Multitasking (MT), Sign Ignore Road (DIT), Tracking Ignore Discrimination (TID) and Ignore All (IA) conditions were specifically interrogated during this experiment. C. Jittered design illustrating the variable EOAs of 0, 300 and 600 ms for the presentation of each sign event with respect to each tracking event were used in all conditions where both stimuli were present (conditions found in 1B). D. Schematic of active and passive task components by condition.

Figure 2. Distraction and Multitasking Performance Costs with Event Onset Asynchrony. Means of response times (RT) in the DIT (blue) condition and MT (green) condition across each EOAs (0, 300, 600 ms) with standard error shown. Mean response times were baselined to the Discrimination Only (DO) condition for each participant to gauge distraction (DIT – DO) and multitasking costs (MT – DO). A 2 X 3 repeated-measures ANOVA with factors of Condition (MT, DIT) and EOA (0, 300, 600 ms) revealed a main effect of condition ($F_{(1,19)} = 95.7$, $p < .001$), EOA ($F_{(2,38)} = 38.7$, $p < .001$), but no interaction ($F_{(1,19)} = 1.2$, $p = 0.33$). Follow-up t-tests indicated that RTs of the discrimination task were slower while multitasking versus distracted-tasking ($t_{(19)} = 3.17$, $p = 0.005$) and that distracted-tasking was slower than the Discrimination Only condition ($t_{(19)} = 7.68$, $p = 3 \times 10^{-7}$) and that multitasking was slower than the Discrimination Only condition ($t_{(19)} = 6.36$, $p = 4.2 \times 10^{-6}$). Responses took longer for both conditions as the EOA approached complete temporal overlap between tasks (MT0 >MT3 >MT6; DIT0 >DIT3 >DIT6; **see Table 1**). Error bars represent standard error means. +p < 0.05 between temporal synchronies.

Figure 3. Attentional Rivalry and Event Onset Asynchrony Influence on P100 amplitude. A. P100 amplitude means for stimuli in the Discrimination Only (DO) condition, Sign Ignore Road (DIT; blue) condition, Multitasking condition (MT; green), Tracking Ignore Discrimination condition (TID; red) condition and Ignore All (IA; black) condition. P100 amplitudes were baselined against our Ignore All (IA) condition where participants ignored the signs and roads. A one-way ANOVA revealed the conditions to have different P100 modulation ($F_{(4,76)} = 10.7$, $p < 6.5 \times 10^{-7}$). B. P100 amplitudes for all four conditions containing EOA (MT, DIT, TID, IA) are baselined against the control condition (IA) for each EOA. A 2 X 2 X 3 repeated-measures ANOVA with factors of for the effect of Stimulus Relevance (SR), Attentional Rivalry (AR) and Event Onset Asynchrony (EOA; 0 ms, 300 ms, and 600 ms) on the selected electrode group of interest revealed a main effect of EOA ($F_{(1,19)} = 15.0$, $p < 1.6 \times 10^{-5}$), but no main effect of Rivalry ($F_{(1,19)} = 0.55$, $p = 0.93$) or Relevance $F_{(1,19)} = 3.5$, $p < 0.078$). 2-way interactions were significant such that AR*EOA ($F_{(2,38)} = 11.3$, $p < 7.3 \times 10^{-4}$), SR*EOA ($F_{(2,38)} = 8.1$, $p < 0.004$) but not AR*SR ($F_{(2,38)} = 0.21$, $p < 0.81$). Pair-wise comparisons of EOA indicated an attenuating effect of decreasing EOA on P100 amplitudes for DIT, such that the closer events were together, the lower the P100 amplitude (DIT0 <DIT300 <DIT600), with similar patterns observed for the multitasking (MT) condition and TID condition

(**see Table 1**). As the plot shows, the interaction is driven by the values at an event onset asynchrony of 0 ms, where TID and DIT exhibited diminished P100 amplitudes compared to baseline. At an EOA of 600 ms, DIT and MT exhibited enhancement of P100 amplitudes compared to baseline. *$p < 0.05$ between conditions, +$p < 0.05$ between temporal synchronies. C. Waveforms for ERPs. (i) Waveforms for single-task conditions (SO, TO) and the control condition (IA) demonstrating no ERP elicited from road events (TO). (ii) MT across all three synchronies (0, 300, 600 ms). (iii) ERPs for all four conditions (MT, IA, TID, DIT) at 0 ms. (iv) ERPs for all four conditions at 600 ms.

# Illustration and Tables

## Table 1 – Summary of Response Cost analysis (p-values shown)

**T-TESTS**

|  | Multitask (MT) | Sign Ignore Road (DIT) |
|---|---|---|
| 0 vs 300 ms | 0.0037 | $7.8 \times 10^{-5}$ |
| 300 vs 600 ms | 0.0035 | 0.062 |
| 0 vs 600 ms | $1.9 \times 10^{-5}$ | $1.2 \times 10^{-6}$ |

|  | 0 ms | 300 ms | 600 ms | ALL EOAs |
|---|---|---|---|---|
| MT vs DIT | $6.7 \times 10^{-7}$ | $4.9 \times 10^{-9}$ | $1.12 \times 10^{-8}$ | 0.005 |

# Table 2 – Summary of P100 analysis (p-values shown)

**A.**

|  | ANOVA (p-values shown) | |
|---|---|---|
|  | 0 ms | 600 ms |
| Stimulus Relevance | 0.97 | **0.006** |
| Attentional Rivalry | **0.029** | 0.125 |
| Stimulus Relevance * Attentional Rivalry | 0.75 | 0.74 |

**B.**

|  | **Attentional Rivalry** | **Stimulus Relevance** | **T-test comparisons (p-values shown)** | |
|---|---|---|---|---|
|  |  |  | 0 ms | 600 ms |
| DIT | **Present** | Attend | **0.027** | 0.297 |
| MT | **Absent** | Attend | | |
| TID | **Present** | Ignore | **0.053** | 0.352 |
| IA | **Absent** | Ignore | | |
| MT | Absent | **Attend** | 0.854 | **0.015** |
| IA | Absent | **Ignore** | | |
| DIT | Present | **Attend** | 0.88 | **0.030** |
| TID | Present | **Ignore** | | |

# Table 3 – Summary of ERP analysis (p-values shown)

A.

| ANOVA | P100 Latency | P100 Amplitude | N200 Latency | N200 Amplitude |
|---|---|---|---|---|
| **EOA** | **0.0054** | **$3.44 \times 10^{-8}$** | 0.27 | **0.0039** |
| **Condition** | 0.27 | 0.94 | **0.0041** | 0.68 |
| **EOA * Condition** | 0.3 | **0.0006** | 0.13 | 0.18 |

B. P100 Amplitude Interaction Analysis for all Conditions

| ANOVA | MT | IA | TID | DIT |
|---|---|---|---|---|
| **EOA** | **$1.1 \times 10^{-3}$** | 0.946 | **$4.6 \times 10^{-4}$** | **$1.3 \times 10^{-8}$** |

**Figure 1**

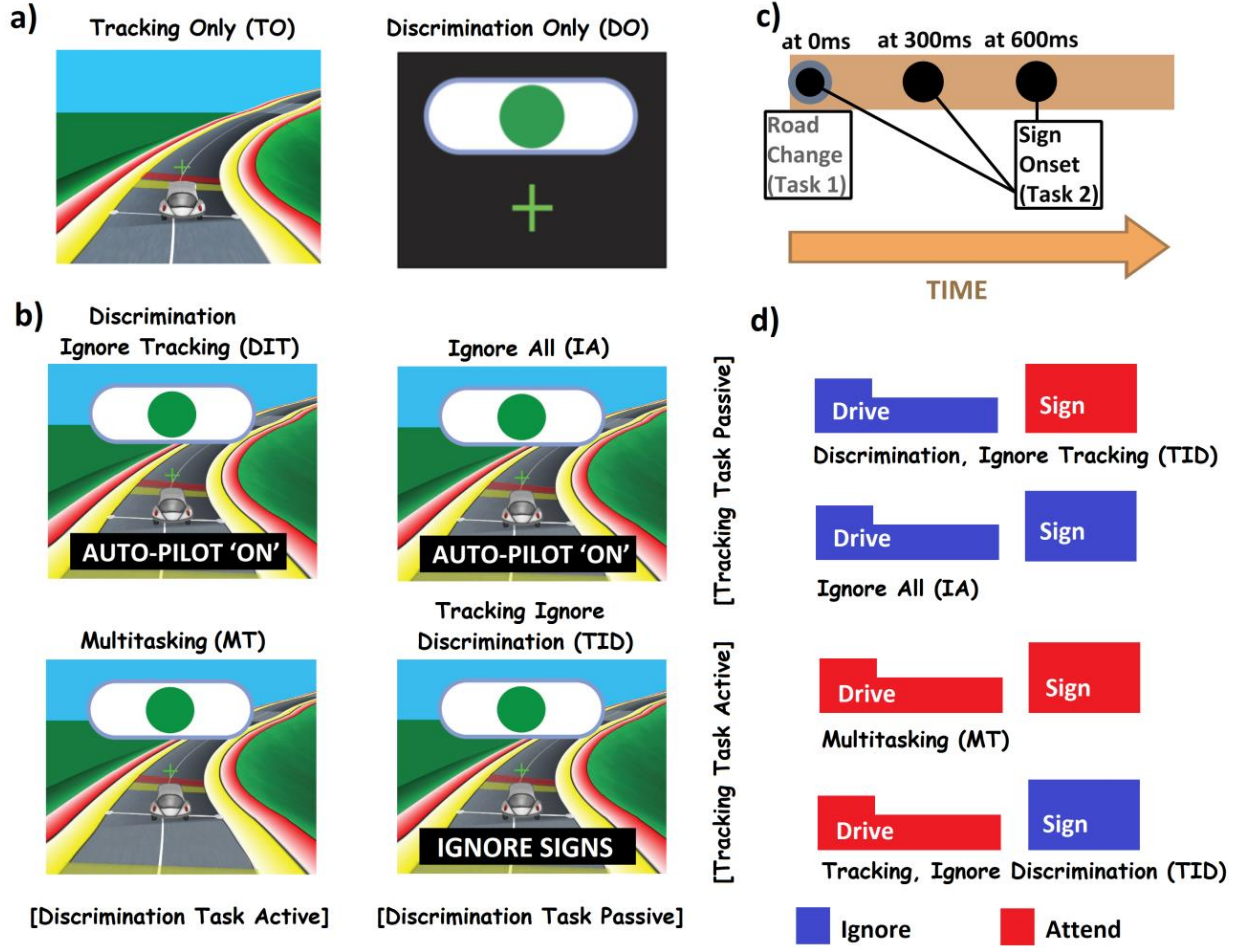

**Figure 2**

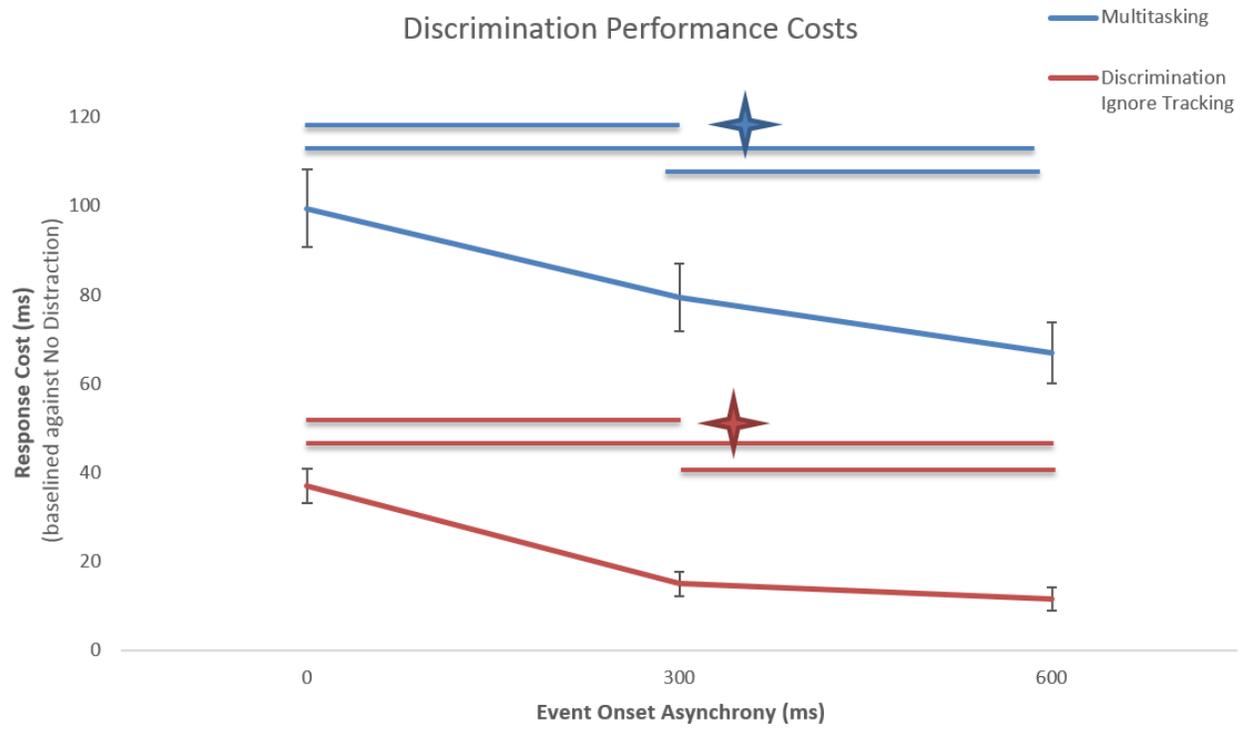

# Figure 3

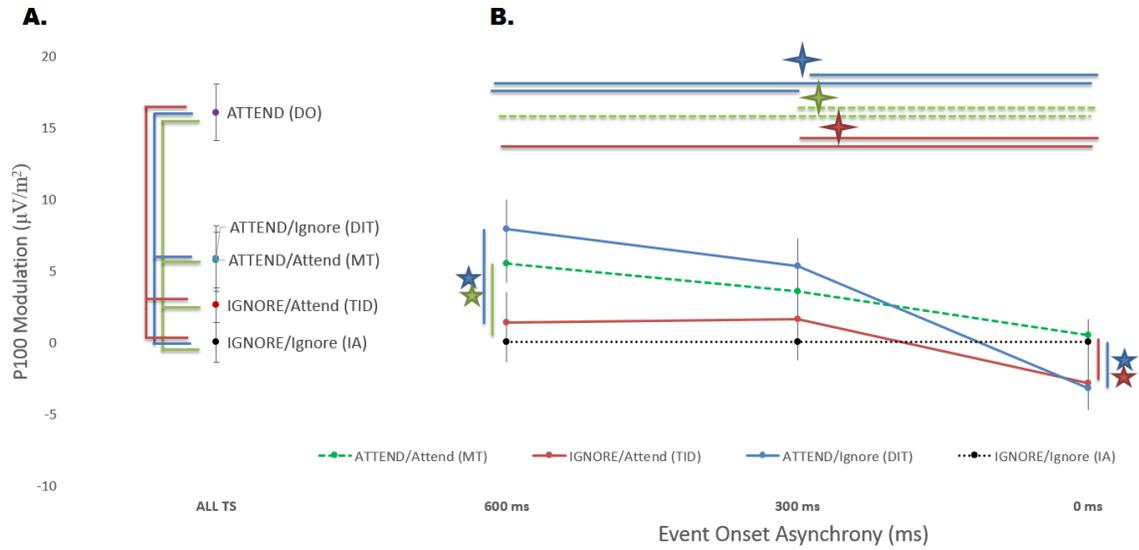

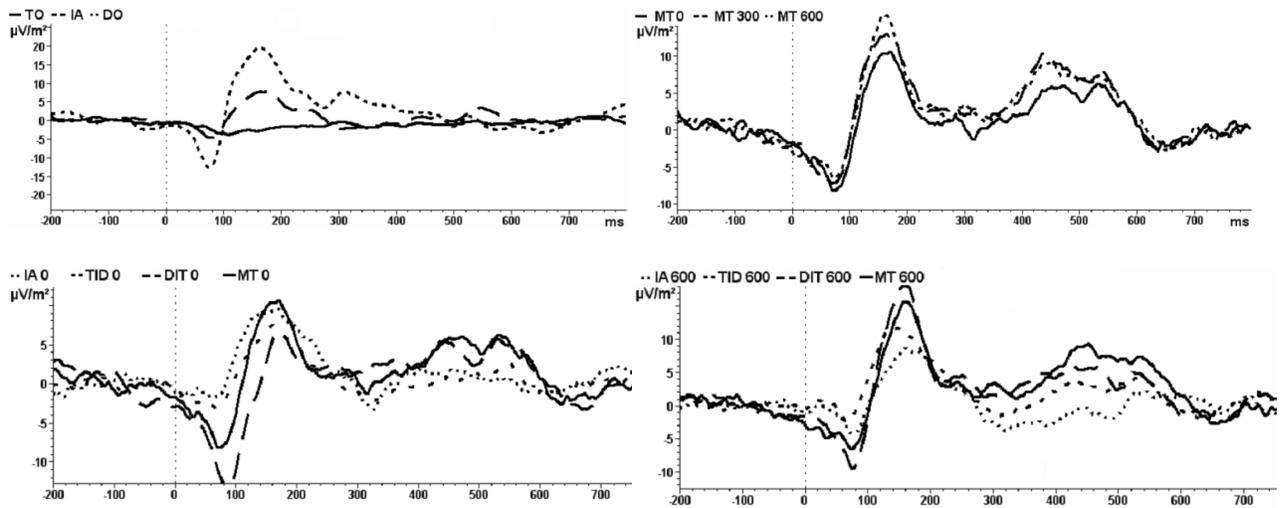